\documentclass[conference]{IEEEtran}

\usepackage{ifpdf}

\usepackage{cite}

\ifCLASSINFOpdf
  \usepackage[pdftex]{graphicx}
\else

\fi

\usepackage{amsmath}
\usepackage[normalem]{ulem}
\usepackage{multirow}
\usepackage{amssymb}
\usepackage{tikz}
\usepackage{lipsum}

\ifCLASSOPTIONcompsoc
  \usepackage[caption=false,font=normalsize,labelfont=sf,textfont=sf]{subfig}
\else
  \usepackage[caption=false,font=footnotesize]{subfig}
\fi
\newcommand\copyrighttext{%
	\footnotesize © 2019 IEEE.Personal use of this material is permitted. Permission from IEEE must be obtained for all other uses, in any current or future media, including reprinting/republishing this material for advertising or promotional purposes,creating new collective works, for resale or redistribution to servers or lists, or reuse of any copyrighted component of this work in other works.}
\newcommand\copyrightnotice{%
	\begin{tikzpicture}[remember picture,overlay]
	\node[anchor=south,yshift=10pt] at (current page.south) {\fbox{\parbox{\dimexpr\textwidth-\fboxsep-\fboxrule\relax}{\copyrighttext}}};
	\end{tikzpicture}%
}

\begin{document}
%
\title{{Nonlinear Dynamical System Model for Drive Mode Amplitude Instabilities in MEMS Gyroscopes}
\\\bigskip \rmfamily\large{\uline{Ulrike Nabholz}\textsuperscript{1}, Michael Curcic\textsuperscript{1}, Jan E. Mehner\textsuperscript{2} and Peter Degenfeld-Schonburg\textsuperscript{1}}  \vspace{0.4cm}\\ {\normalsize \textsuperscript{1}Department of Microsystems and Nanotechnologies, Corporate Research, Robert Bosch GmbH, Renningen, Germany
\\\textsuperscript{2}Faculty of Electrical Engineering and Information Technology, Chemnitz University of Technology, Chemnitz, Germany}\vspace{-0.8cm}}

\maketitle

\begin{abstract}
The requirements pertaining to the reliability and accuracy of micro-electromechanical gyroscopic sensors are increasing, as systems for vehicle localization emerge as an enabling factor for autonomous driving. Since micro-electromechanical systems (MEMS) became a mature technology, the modelling techniques used for predicting their behaviour expanded from mostly linear approaches to include nonlinear dynamic effects. This leads to an increased understanding of the various nonlinear phenomena that limit the performance of MEMS sensors. In this work, we develop a model of two nonlinearly coupled mechanical modes and employ it to explain measured drive mode instabilities in MEMS gyroscopes. Due to 3:1 internal resonance between the drive mode and a parasitic mode, energy transfer within the conservative system occurs. From measurements of amplitude response curves showing hysteresis effects, we extract all nonlinear system parameters and conclude that the steady-state model needs to be expanded by a transient simulation in order to fully explain the measured system behaviour.\\
\end{abstract}
\copyrightnotice

\section{Introduction}
For automotive and consumer applications, where size, cost and reliability are the main concerns, gyroscopes are typically designed as micro-electromechanical systems (MEMS) \cite{Neul2007}. They can be affected by adverse effects, most importantly instabilities in the zero-rate offset values \cite{Saukoski2007}. In mode-matched gyroscopes, two modes of oscillation of the same frequency, drive mode and detection mode, perform vibrating motions in orthogonal directions. Within the scope of this paper, we focus on the influence of higher-frequency modes of oscillation on the drive mode. \\
Coupled oscillations, often nonlinear in nature, have been studied extensively, and a range of analytical as well as numerical reduced-order modelling approaches were proposed \cite{Nayfeh2008, Strogatz2007}. The application of nonlinear mode coupling of vibrational modes in micro-electromechanical systems has been studied under various regimes of internal resonance \cite{Ganesan2017, Phani2006, Venstra2012, Samanta2015}, and has also been applied to MEMS gyroscopes \cite{Putnik2016}.\\
Our modelling approach introduced in \cite{Nabholz2018_arxiv} is based on a strain energy formulation of the elastic strain energy and thus applicable to any oscillating system driven at resonance. Here, measurements suggest the relevant nonlinear terms which can be modelled using a two degree-of-freedom (DOF) system for the drive mode and a parasitic mode.
\section{Measurements}
\begin{figure}[ht]
	\centering
	\includegraphics[width=0.8\linewidth]{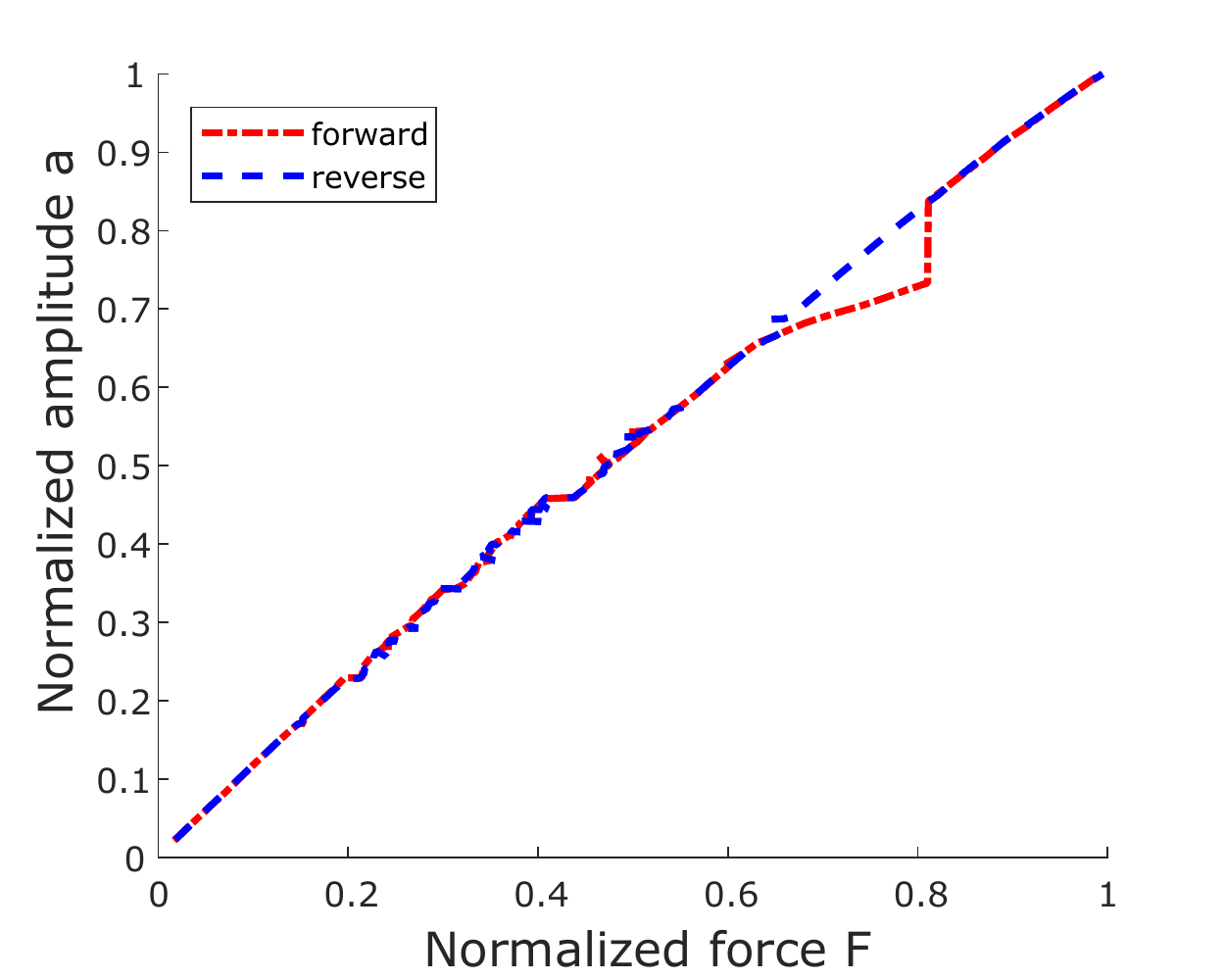}
	\caption{Measurements of the device operated with a phase-locked loop (PLL) set to \(\phi_0 = -\frac{\pi}{2}\). Deviations between forward (red) and reverse (blue) sweeps showcase hysteresis effects that lead to amplitude instabilites and jump phenomena in the drive mode.}
	\label{fig:measured_pll}	
\end{figure}

\begin{figure*}[h]
	\centering
	\includegraphics[width=0.85\textwidth]{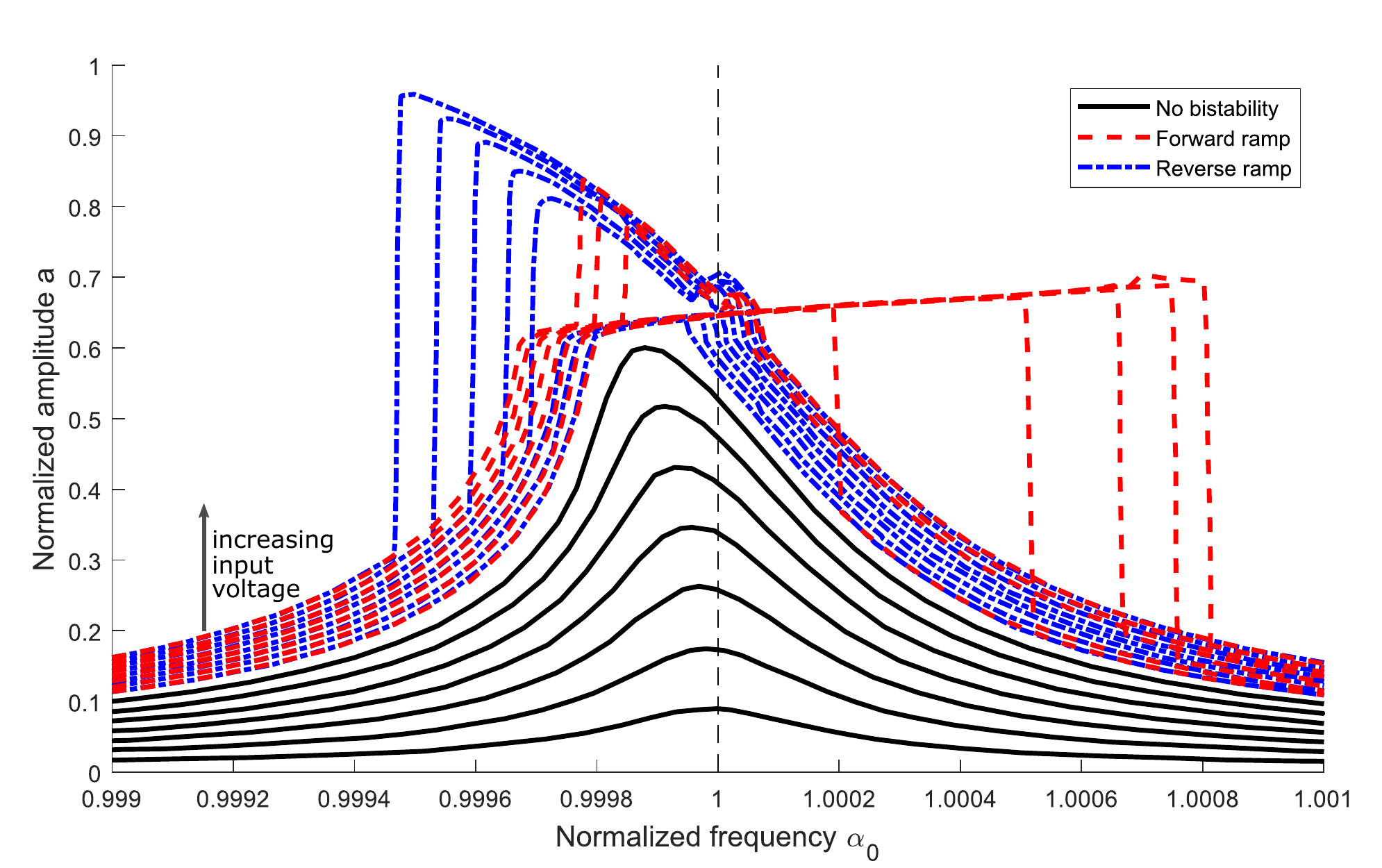}
	\caption{Measured amplitude response functions for different input voltages. Black lines denote identical forward and reverse frequency sweeps for the lower voltages. For higher voltages, forward and reverse sweep amplitudes are different; red lines now denote the forward sweep, blue lines the reverse sweep. The instability visible at \(\alpha_0 \approx 1\) for high voltages indicates dynamic behaviour.}
	\label{fig:1}
\end{figure*}

\noindent
Using a phase-locked loop (PLL), where the phase of the drive frequency was set to \(\phi_0 = -\frac{\pi}{2}\), the behaviour of a gyroscopic sensor was characterized. The results are shown in Fig.\,\ref{fig:measured_pll}. 
An electrical characterization method with a carrier signal \cite{Cigada2007} was used to investigate the mechanical behaviour of an unpackaged sensor.\\
The resulting amplitude response curves for a range of different input voltages are shown in Fig.\,\ref{fig:1}. Apparently, for higher amplitudes, bistable behaviour and jump phenomena occur and thus, forward and reverse sweeps differ and even exhibit crossings.\\
\noindent 

\section{System Model}
From nonlinear mechanics \cite{Landau1986}, we derive a strain energy formulation in modal form with three- and four-wave terms using the Green-Lagrange strain measure, as introduced previously \cite{Nabholz2018_arxiv}.
When deriving our coupled equation of motion from the strain energy, we can omit all non-resonant terms, since we are limiting our analysis to high quality factor MEMS.
An analysis of the measured frequency spectrum of the device suggests that a mode at roughly triple the drive frequency gains amplitude. At the same time, the amplitude of the drive mode is lower than expected. This points to the occurrence of 3:1 internal resonance and we can thus deduce the following equations of motion for our two degree-of-freedom system from the modal strain energy:
%
%
\begin{IEEEeqnarray}{rcl}
	\ddot{q}_{0}\! + \!\left(\omega_{0}^{2} \!+\! \beta_{0} q_{0}^{2}\! +\! V q_{1}^{2} \right)\! q_{0} \! +\! \frac{\omega_{0}}{Q_{0}} \dot{q}_{0}\! +\! 3 \chi q_0^2 q_1 & = & F_{0} \sin\!\left(\omega_{d} t\right)\!, \label{eom_mode1} \\
	\ddot{q}_{1}\! +\! \left(\omega_{1}^{2}\! +\! \beta_{1} q_{1}^{2}\! +\! V q_{0}^{2} \right)\! q_{1}\! +\! \frac{\omega_{1}}{Q_{1}}\dot{q}_{1}\! +\! \chi q_0^3 & = & 0. \label{eom_mode2} 
\end{IEEEeqnarray}
The model parameters and all the parameters used so far as well as used in the following derivation of our system model are given in Table \ref{tab:parameters}. Note that all time-dependencies of the modal amplitudes \(q_i\left(t\right)\) with \(i = 0,1\) have been omitted to enhance readability.\\
\begin{table}[ht]
	\renewcommand{\arraystretch}{1.4}
	\centering
	\begin{tabular}{|c | c | c |}
		\hline
		\textbf{Parameter} & \multicolumn{2}{c|}{\textbf{Description}}\\
		\hline
		\(q_i\) & \multirow{4}{*}{Amplitude} &  Modal amplitude of mode i  \\
		\(a_i\) & & Amplitude of mode i \\
		\(a = \frac{a_0}{a_{0,max}}\) & & Normalized drive mode amp. \\
		\hline
		\(f_{i}\) & \multirow{5}{*}{Frequency} & Frequency of mode i \\
		\(f_{d}\) & & Drive frequency \\
		\(\omega_{i}\) & & Angular mode freq. of mode i \\
		\(\omega_d\) & & Angular drive frequency \\
		\(\alpha_i = \frac{f_{d}}{f_{i}}\) & & Normalized freq. of mode i \\
		\hline
		\(\beta_i\) & \multirow{3}{*}{Nonlinear Coeff.} & Duffing coefficient of mode i \\
		\(V\) & & Cross-Duffing coefficient \\
		\(\chi\) & & Upconversion coefficient \\
		\hline
		\(Q_i\) & \multirow{6}{*}{Miscellaneous} & Quality factor of mode i \\
		\(d\) & & Detuning \\
		\(F_0\)  & & Input force amplitude \\
		\(F = \frac{F_0}{F_{0,max}}\) & & Normalized force \\
		\(t\) & & Real time scale \\
		\(\tau = t\,\omega_{0}\) & & Slow time scale \\
		\hline
		
	\end{tabular}
	\vspace{0.2cm}
	\label{tab:parameters}
	\caption{Model Parameters}
\end{table}
Here, the index 0 denotes the drive mode oscillation, whereas the index 1 denotes the parasitic mode. The detuning \(d\) establishes a relation between the two linear mode frequencies and is dependent on the distribution of the linear mode frequencies in the specific device. It denotes the deviation of the frequency of the parasitic mode from the multiple of the driving frequency:
\begin{equation}
\label{omega1}
\omega_{0,1} = n\cdot \omega_{0,0} + 2\pi\cdot d \approx n \cdot \omega_{0,0} \qquad \text{ for n } \in \mathbb{N}, n>1 .
\end{equation}
In the case of 3:1 internal resonance, we assume \(n = 3\).\\
Our model comprises four nonlinear coefficients: The Duffing coefficient \(\beta_i\) of each mode \(i\), the mutual frequency shift between the two modes which we call Cross-Duffing coefficient \(V\), and the coefficient for 3:1 internal resonance \cite{Nayfeh2008} termed the upconversion coefficient \(\chi\). The nonlinear coefficients and the quality factors are extracted from the measurements shown in Fig. \ref{fig:1}. \(F_0\) denotes the amplitude of the external periodic force that actuates the drive mode.\\
%
%
\noindent 
We employ the method of averaging \cite{Bogoliubov1961} to reduce the system to first order differential equations for amplitude and phase of each mode, as shown previously \cite{Nabholz2018}. This approach is valid for resonant systems with high quality factors, where two time scales for fast and slow oscillation can be identified. Assuming steady-state, we obtain implicit analytical equations for all solution branches:
The method of averaging for the additional terms is carried out as shown by \cite{Nabholz2018} and yields for \(n=3\)
\begin{IEEEeqnarray}{rcl}
	\dot{a}_{0}\left(\tau\right) & \: = \: & -\frac{1}{2 Q_{0}} a_{0}\left(\tau\right) - \frac{F_{0}}{2\alpha_0\omega_{0}^2} sin\left(\phi_{0}\left(\tau\right)\right) \nonumber \\
	& \: - \: & \frac{3\chi}{8\alpha_0\omega_0^2}a_0^2\left(\tau\right) a_1\left(\tau\right) \sin\left(3\phi_0\left(\tau\right)-\phi_1\left(\tau\right)\right) \label{1.1} \\
	\dot{\phi}_{0}\left(\tau\right) & \: = \: & \frac{\sigma_{0}}{2\alpha_{0}} + \frac{3\beta_{0}}{8\alpha_0\omega_{0}^2} a_{0}^2\left(\tau\right) + \frac{V}{4\alpha_0\omega_{0}^2}a_{1}^2\left(\tau\right)\nonumber \\
	& \:-\: & \frac{F_{0}}{2\alpha_0\omega_{0}^2 a_{0}\left(\tau\right)}\cos\left(\phi_{0}\left(\tau\right)\right) \nonumber \\
	&&- \frac{3\chi}{8\alpha_0\omega_0^2} a_0\left(\tau\right) a_1\left(\tau\right) \cos\left(3\phi_0\left(\tau\right)-\phi_1\left(\tau\right)\right), \label{1.2} \\
	\dot{a}_{1}\left(\tau\right) & \: = \: & -\frac{\omega_{1}}{2 Q_{1}\omega_{0}}a_{1}\left(\tau\right) \nonumber \\
	&&+ \frac{\chi}{8\alpha_1\omega_0^2} a_0^3\left(\tau\right) \sin\left(3\phi_0\left(\tau\right)-\phi_1\left(\tau\right)\right), \label{2.1} \\
	\dot{\phi}_{1}\left(\tau\right) & \: = \: & \frac{\sigma_{1}}{2 \alpha_1} + \frac{3\beta_{1}}{8 \alpha_1 \omega_{0}^2} a_{1}^2\left(\tau\right) + \frac{V}{4 \alpha_{1} \omega_{0}^2} a_{0}^2\left(\tau\right) \nonumber \\
	&&- \frac{\chi}{8\alpha_1\omega_0^2 a_1\left(\tau\right)} a_0^3\left(\tau\right) \cos\left(3\phi_0\left(\tau\right)-\phi_1\left(\tau\right)\right). \label{2.2}  
\end{IEEEeqnarray}
It has to be noted that the upconversion terms only contribute to the steady-state equation, when \(n\) is a multiple of 3. For other values of \(n\), the averaging procedure eliminates the terms in both amplitude and phase equations.
%
%
%
\section{Results}
\begin{figure}[ht]
	\centering
	\includegraphics[width=0.75\linewidth]{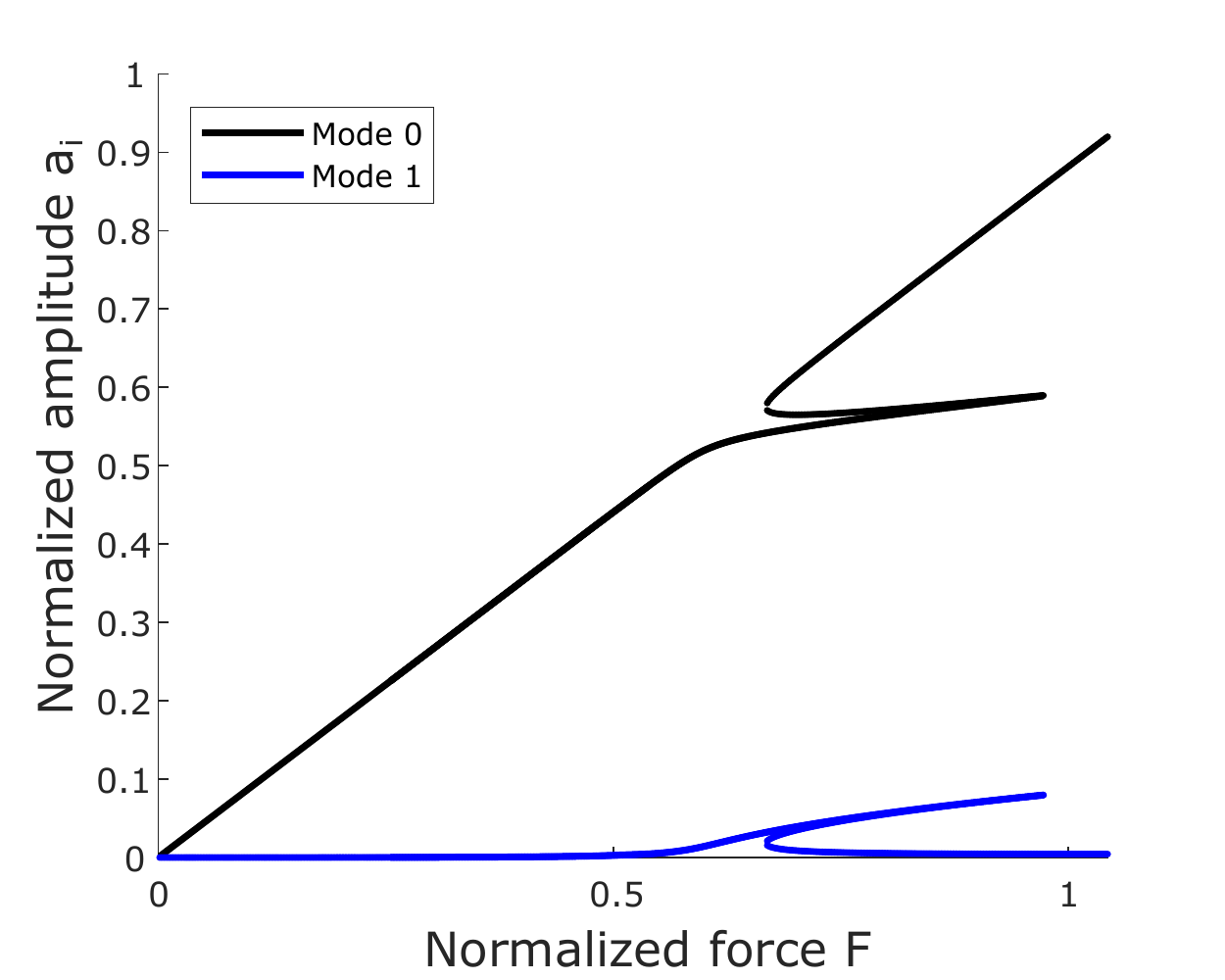}
	\label{fig:modelled_pll}
	\caption{Modelled two-mode system set to the same phase relation as used in the measurements in Fig.\,\ref{fig:measured_pll}. The internal resonance model can explain the hysteretic behaviour observed in the measurements with the energy transfer that occurs and depletes the amplitude of mode 0}
\end{figure}
Setting the phase of the drive mode \(\phi_0 = -\frac{\pi}{2}\) in the steady-state equations (\ref{1.1}-\ref{2.2}) yields the results shown in Fig.\,\ref{fig:modelled_pll}. A comparison with the PLL measurements in Fig.\,\ref{fig:measured_pll} showcases the energy transfer that occurs between drive mode and parasitic mode in the region of hysteresis.\\
\begin{figure}[t]
	\centering
	\subfloat[Results of the forward frequency sweep (blue) superimposed onto the steady-state solutions and the measured amplitude response curve in forward direction (red).]
	{
		\includegraphics[width=0.95\linewidth]{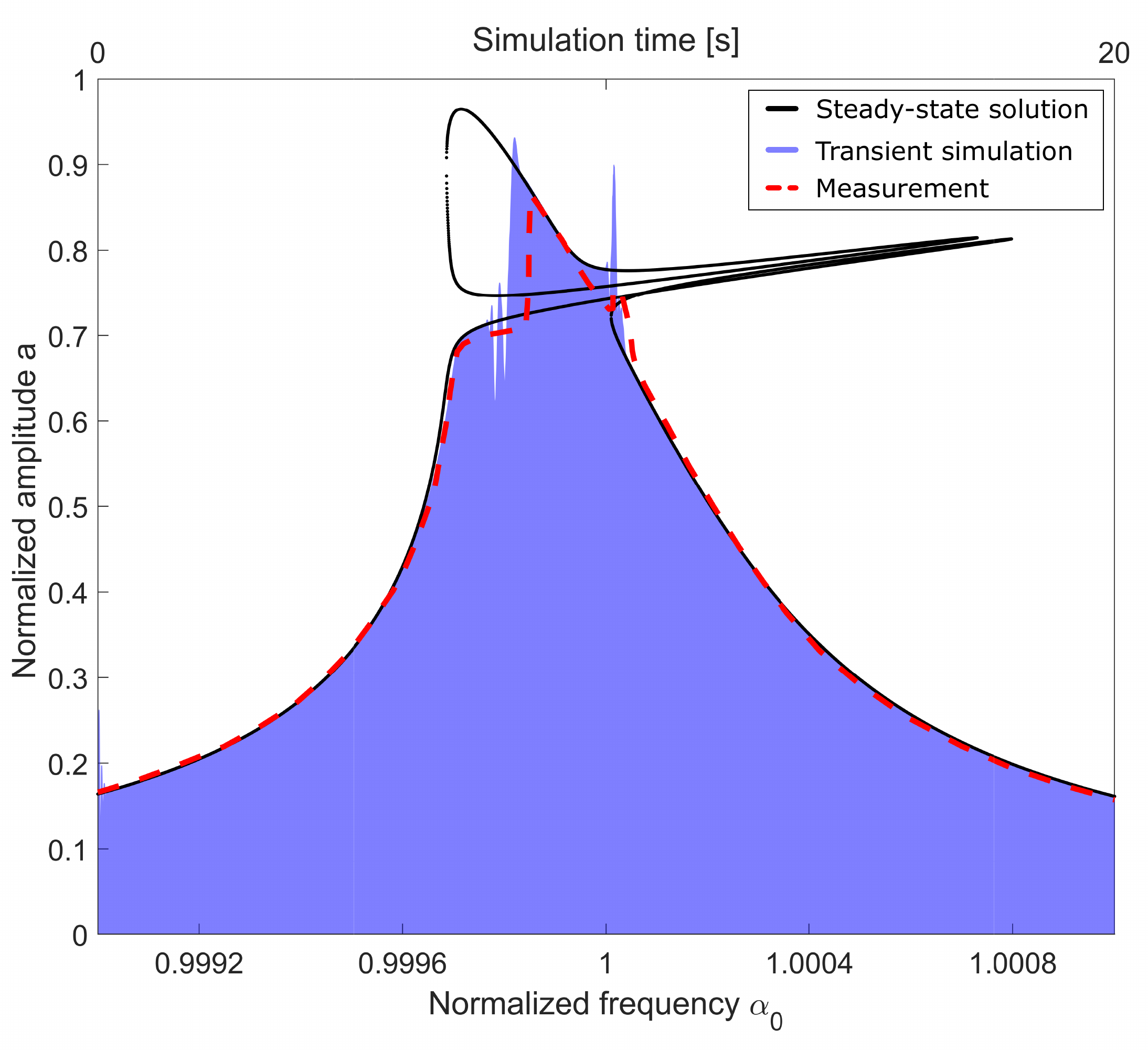}
		\label{SUBFIGURE:modelled_up}
	}
	\hfil
	\subfloat[Results of the reverse frequency sweep (blue) superimposed onto the steady-state solutions and the measured amplitude response curve in reverse direction (red). ]
	{
		\includegraphics[width=0.95\linewidth]{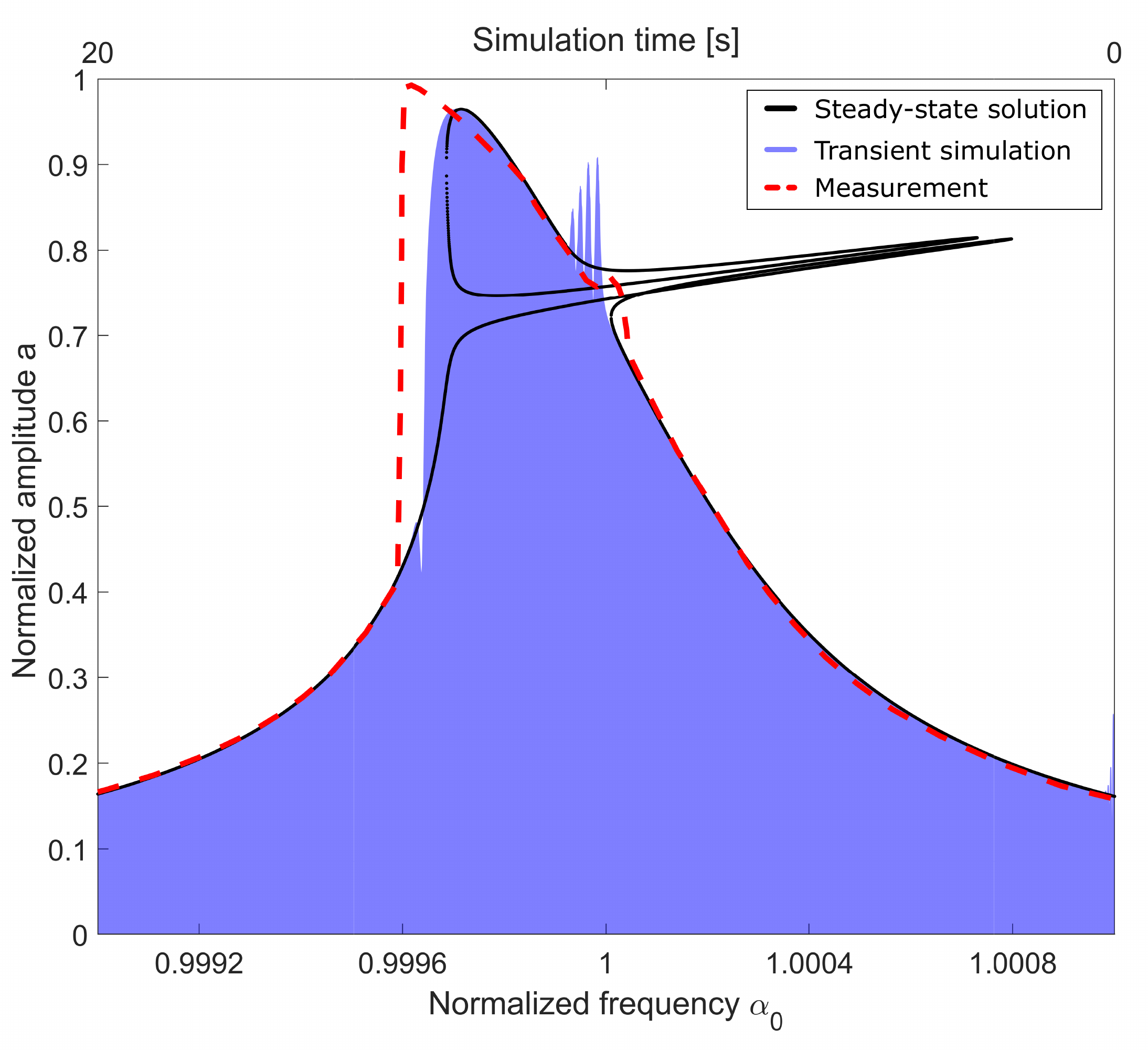}
		\label{SUBFIGURE:modelled_down}
	}
	\caption{Modelled amplitude response curves for an exemplary input voltage: The transient simulation results for forward or reverse frequency sweeps are plotted as shaded areas superimposed onto the steady-state solutions. The lower horizontal axis shows the covered frequency range given by the normalized drive frequency \(\alpha_0\), the upper horizontal axis the corresponding simulation time. The vertical axis shows the normalized amplitude of the oscillation.}
	\label{fig:2}	
\end{figure}
\noindent
The steady-state results of equations (\ref{1.1}-\ref{2.2}) without a fixed phase-relation simulated for an exemplary input force, shown in Fig.\,\ref{fig:2}, denote all solution branches obtained, independent of their stability. This leads to the question of how to predict the behaviour of the real system, i.e. which solution branch is actually reached during operation. If we maintain the assumption that the system settles into steady-state for any possible parameter combination, we cannot explain the transitions between solution branches that occur in both forward and reverse measurements. However, Fig. \ref{fig:1} shows a bump in the reverse frequency sweeps at around \(\alpha = 1\). Thus, we suspect that dynamic effects such as limit cycles play an important role in the transition to the upper solution branch. \\
We also compare the measured frequency sweeps for various input forces \(F_0\) in Fig.\,\ref{fig:1} with the exemplary steady-state solutions in Fig.\,\ref{fig:2}. This clearly shows that the solution branches that resemble an amplitude plateau towards higher frequencies and are reached in measurements for medium high input forces, do not vanish in the modelled system, but rather become unstable as the input force is increased. A stability analysis shows that some solution branches indeed change stability as the input force \(F_0\) is varied. \\
In order to simulate the influence of the unstable regions onto the system behaviour, we drop the steady-state assumption and conduct transient frequency sweeps of the two coupled equations of motion: Shown in Fig.\,\ref{fig:2}, transitions between solution branches occur and we observe dynamic system behaviour in the form of large amplitude modulations in these transition regions. Especially the forward sweep in Fig.\,\ref{SUBFIGURE:modelled_up} shows how the transition from lower to upper solution branch occurs as the lower branch becomes unstable. Both forward and reverse sweep show the expected limit cycle behaviour around \(\alpha = 1\), confirming the absence of a single stable solution branch.\\
With this transient expansion of our simulation, we can successfully emulate our measurements. 

\noindent
%

%
%
%
\section{Conclusion}
We showed the validity of our simple two degree-of-freedom system model for emulating measurements of amplitude response curves that show instabilities of the drive mode: The analytic steady-state equations yield all possible solution branches, whereas parameter sweeps of the transient system explain the dynamic effects. Since our model is confined to only the relevant terms and modes of oscillation, the required computational effort is very small. \\
Thus, our approach allows us to understand and characterize complex MEMS gyroscopes, leading to a prediction for future designs. This forms a basis for redesigns eventually leading to a stable design with distributions of parasitic modes such that the drive mode is not disturbed.\\
In general, the variability of MEMS process technologies leads to a normal distribution of the system's linear mode frequencies and thus, a wide range of mode couplings is possible. Through design modifications, the frequency of a mode can be moved away from multiples of another mode, yet the absolute number of possible mode couplings is so large that not all of them can be prevented. The effect analysed in this paper occurs only in very few manufactured devices, since the strength of the nonlinear coupling largely depends on how closely the two modes fulfil the frequency condition for 3:1 internal resonance, i.e. on the amount of detuning \(d\).\\
As a first step, we aim to assess the qualitative influence of each system parameter on the amplitude response curve and thus on the mode coupling behaviour. With this knowledge and by conducting further measurements of test devices manufactured specifically to meet the frequency ranges of interest, we aim to establish design rules using correlations between coupling coefficients and design geometry. \\

\bibliography{abstract_bibliography}
\bibliographystyle{ieeetr}

\end{document}